\documentclass[twocolumn]{webofc}

\usepackage[varg]{txfonts}   
\begin{document}
\title{Toward a quantum-enhanced strontium optical lattice clock at INRIM}
%
%

\author{\firstname{Marco G.} \lastname{Tarallo}\inst{1}\fnsep\thanks{\email{m.tarallo@inrim.it}}}

\institute{Istituto Nazionale di Ricerca Metrologica, Strada delle Cacce 91, 10135 Torino (Italy) 
          }

\abstract{%
  The new strontium atomic clock at INRIM seeks to establish a new frontier in quantum measurement by joining state-of-the-art optical lattice clocks and the quantized electromagnetic field provided by a cavity QED setup. The goal of our experiment is to apply advanced quantum techniques to state-of-the-art optical lattice clocks, demonstrating enhanced sensitivity while preserving long coherence times and the highest accuracy. In this paper we describe the current status of the experiment and the prospected sensitivity gain for the designed cavity QED setup.
}
\maketitle
\section{Introduction}
\label{intro}

The optical lattice clock (OLC)~\cite{takamoto2005,ludlow2015} is nowadays the most sensitive~\cite{oelker2019} and precise~\cite{mcgrew2018} device for the measurement of time and frequency, reaching the $1\times10^{-18}$ uncertainty level with instabilities below $10^{-16}$ at 1 s. The range of potential applications of such incredible precision spans from improved satellite navigation and deep space missions, to telecommunications, stock exchange transaction, and very-long baseline interferometry~\cite{clivati2017}. It is also a powerful device to perform more and more stringent tests of fundamental physics~\cite{mcgrew2019,wcislo2018}, in particular on-field tests of General Relativity which may pave the way to relativistic geodesy~\cite{grotti2018}.  

The recipe for an optical lattice clock is still a ``classical'' one. Its ingredients are: i) $N$ identical and independent frequency discriminators, which are provided by laser-cooled atoms optically-trapped in such a way that their motion along the probing direction is fully quantized and thus motional systematic effects are suppressed; ii) a local oscillator, which is provided by a laser field already stabilized to an ultra-stable optical resonator; iii) a counting device, which is provided by the optical frequency comb. The clock is operated in a pulsed sequence of duration $T_c$, in which nearly half of the time is spent for the preparation of the cold atomic sample and the other half consists of the clock laser-atom interaction. Finally, the excitation fraction $P_e$ is estimated by a destructive projective measurement. In order to increase the clock precision, averaging is usually performed by repeating this sequence for a longer time $\tau$. The ultimate limit to the instability of an optical lattice clock, operated on an optical transition with a quality factor $Q=\omega_0/\Delta\omega$, is set by the probabilistic nature of the projective measurement outcome, which for a set of $N$ uncorrelated atoms is the so-called standard quantum limit~\cite{hollberg2001}

\begin{equation}\label{eq:sql}
    \sigma_y(\tau) \approx \frac{1}{\pi Q\sqrt{N}}\sqrt{\frac{T_c}{\tau}}\,.
\end{equation}

The pulsed operation of the optical lattice clock affects its stability by means of the so-called Dick effect~\cite{westergaard2010}, for which laser frequency noise at Fourier frequencies higher than the clock cycle frequency are down-converted by aliasing to an effective carrier frequency shift and then translated into frequency instability by the clock servo loop. The Dick effect is currently the main limitation to the optical lattice clock stability.

While several ``classical'' solutions are currently investigated to overcome the stability limitations, such as cryogenic optical resonators with intrinsic lower frequency noise~\cite{matei2017}, synchronously operated pairs of clocks~\cite{schioppo2017,hume2016}, or clock operation in optical tweezers with reduced dead-time~\cite{norcia2019}, it is possible to employ quantum technologies to enhance the stability of optical lattice clocks, re-thinking both the measurement protocols and the local oscillator laser generation. In this paper we describe the approach undertaken at INRIM based on a cavity-enhanced system which allows to study both non-classical collective states with quantum projection noise lower than the standard quantum limit, and the generation of highly coherent laser radiation by superradiant atomic dipoles collectively strongly-coupled to an optical cavity. In particular, in Sec.~\ref{sec-1} we describe how quantum technologies can be employed to improve the optical lattice clock. Section~\ref{sec:3} details the design of the cavity-enhanced strontium optical clock at INRIM. Finally, in Sec.~\ref{sec:4} we show the current progress of our experiment, in particular the sideband-enhanced 2D-MOT atomic source we recently developed~\cite{barbiero2019}.

\section{Quantum-enhancement in optical lattice clocks}
\label{sec-1}

The optical lattice clock, as any atomic clock, is a device which generates and/or steers a generated/external optical frequency $\omega_L$ with respect to the chosen atomic energy transition with frequency $\omega_0$ by means of the Rabi flopping mechanism. Here a local oscillator frequency can be encoded in an atomic phase which can be precisely detected by projective measurements on the atomic level populations, or it can trigger the stimulated emission on the clock transition itself, producing an active frequency standard whose instability is governed by the quantum dynamics of the collective atomic system.

From a quantum-mechanical point of view, the passive clock system gives rise to a parameter estimation problem, which can be optimized according to quantum information theory. The parameter estimation problem for atomic clocks was first studied by Wineland et al.~\cite{Wineland1992,Itano1993,Wineland1994} in terms of a collective pseudo-spin operator $\hat{\vec{J}}=\sum^N_i\frac{1}{2}\vec{\sigma}_i$, where the Pauli matrices $\vec{\sigma}_i$ refer to the individual atom two-level pseudo-spin, and the projection operator to the $z$-axis quantifies the population inversion. A general approach to evaluate the maximum achievable precision in atomic phase estimation $\phi = (\omega_0-\omega_L)T_R$, where $T_R$ is the Ramsey precession period, consists on studying the quantum Fisher information $F_Q[\hat{\rho}_\phi]$ of the collective state $\hat{\rho}_\phi$. The quantum Fisher information quantifies the information about an unknown parameter $\phi$ that can be extracted from measurements of a quantum operator (i.e., in the classical version, a random variable), in our case a collective spin projection. The maximum of the quantum Fisher information sets the limit on maximum achievable precision, the so-called quantum Cramer-Rao bound (QCRB) $\Delta\theta_{QCR} = 1/\sqrt{ F_Q[\hat{\rho}_\phi]}\leq 1/N$~\cite{Pezze2009}.

In the case of $N$ uncorrelated atoms exhibiting the same response to unitary transformations in the Bloch sphere, i.e. the so-called coherent spin state (CSS), the limit to frequency estimation is 

\begin{equation}
\Delta\omega = \Delta J_z \left(\frac{\partial J_z}{\partial\omega}\right)^{-1}\bigg|_{J_z = 0} \propto \frac{\Delta J_z^{CSS}}{\langle J\rangle} =\frac{1}{\sqrt{N}}
\end{equation}
which leads to Eq.\ref{eq:sql} and corresponds to the classical error when $N$ uncorrelated systems are probed. The coherent spin state is a minimum uncertainty state because it saturates the Heisenberg relation $\Delta J_z \Delta J_y \geq|J_x|/2$. In terms of QCRB, the CSS has a QCRB = $N$. A QCRB $>N$ implies particle entanglement, up to the $N$ particle entanglement case which leads to a maximum QCRB $=N^2$.

A possible choice for better clock frequency estimation is choosing a state with a reduced $\Delta J_z$ uncertainty

$$
(\Delta J_z^s)^2 =\xi_R^2 (\Delta J_z^{CSS})^2 \frac{\langle J_s\rangle^2}{N^2/4}= \xi_R^2 \frac{\langle J_s\rangle^2}{N}
$$
The parameter $\xi_R^{-2}$ quantifies the metrological gain of a collective state with ``squeezed'' uncertainty orthogonal to the measurement direction weighted by the contrast reduction ~\cite{Wineland1994,Wineland1992}. These so-called squeezed states have the potential to achieve a phase sensitivity proportional to $1/\sqrt{QCRB} =N^{-5/6}$. 

The creation of collective states with metrologically useful entanglement requires a high degree of quantum control with very stringent properties, and even more for the application to state-of-the-art optical clocks. In fact, for a quantum-enhanced optical clock one must require:
\begin{itemize}
    \item optimal control over the atomic state with minimal perturbation (decoherence) of the system;
    \item enhanced and/or tunable coupling strength;
    \item preferably simple system modeling.
\end{itemize}
One also wants to keep the metrological advantage of the OLC system, i.e., keep probing the atoms in the Lamb-Dicke regime~\cite{ludlow2015} in an AC Stark shift free lattice, enabling long ($\sim$ 1 s) coherence time. The latter excludes the employment of engineered collisional interactions to generate entangled states. In this perspective, optical interfaces present all the above prescribed properties and little and controllable perturbation to the atomic reference system. Indeed, looking at the best achieved spin-squeezing results, we can notice that the highest metrological gain factors have been obtained by atom-cavity coupled systems~\cite{Hosten16,Pezze18}.

Collective atomic states can be also created and exploited for direct generation of phase-coherent radiation. In the atom-cavity system, their coupling is a competition between coherent exchange of excitations, and loss of excitations through decay through atomic free space scattering $\Gamma$ and spontaneous decay through both cavity mirrors $\kappa$. The enhanced atom-cavity coupling rate arises from a constructive interference for emission and absorption with respect to the cavity mode due to correlations among the individual atoms. This constructive interference among the individual atomic dipoles may result in an enhancement by $N$ of the excited state decay rate, and thus the total power emitted by the ensemble. This phenomenon is called Dicke superradiance~\cite{Dicke54}. Continuously repumping atoms to the excited state of the clock transition can even enable the construction of continuously operated superradiant lasers~\cite{Meiser2010,bohnet2012}, realizing an optical version of the hydrogen maser.

Unlike ordinary lasers where the coherence of the system is defined by the cavity decay rate rather than the gain medium (i.e., $\kappa\ll\Gamma_{\perp}$), the superradiant laser is operated in the so-called ``bad cavity'' regime, i.e., $\kappa\gg\Gamma_{\perp}$, so that the atoms are the primary carriers of phase information. The superradiant laser has the potential to exploit narrow linewidth transitions, such as the clock transition in an alkaline-earth atom, to generate extreme spectrally-narrow radiation~\cite{Meiser2010}

$$
\Delta\nu_{SR} \geq \frac{C_0\Gamma_\perp}{\pi},
$$
where $C_0$ is the single atom-cavity cooperativity parameter~\cite{Tanji2011}, and $\Gamma_\perp = \Gamma/2+1/T_2$ is the transverse decoherence rate due to atomic spontaneous emission $\Gamma$ and other atomic dephasing mechanism with characteristic time $T_2$. Lowering the cavity cooperativity would reduce the superradiant laser linewidth, but also the total emitted power $P_{SR} \leq \hbar\omega N^2 C_0 \Gamma/8$. Then, near unity cooperativity on the Sr clock transition would yield both spectrally-narrow linewidth and sufficient power to stabilize a bright 1 Hz linewidth laser down to mHz level and perform clock spectroscopy below the standard quantum limit~\cite{bohnet2012}.

All the above considerations brought us to design our new Sr clock apparatus as a merge of an optical lattice clock and a cavity quantum electrodynamics experiment (cQED)~\cite{Walther2006}, having an optical cavity in the regime of collective strong-coupling regime. cQED enjoys all the design requirements we looked for a quantum-enhanced optical clock: light-atom coupling enhanced by the cavity finesse; ii)  extremely low-noise readout techniques are available; only a single-mode photonic field, which is modeled by the well-known Jaynes-Cummings model. Collective strong coupling indicates that, even if the single atom-photon excitation exchange rate $g$ is lower than the system decay rates, their collective spin $\hat{J}$, having its coupling strength enhanced by $\sqrt{N}$, is strongly coupled to the cavity mode. This regime has been already observed for Sr atoms on the two weak intercombination transitions~\cite{Norcia16}. It is also the natural testbed to study superradiant laser generation. The experimental features of INRIM's cavity enhanced Sr optical clock are detailed in the next section.





\section{Design of the cavity-enhanced Sr optical clock}\label{sec:3}

We have designed a cavity-enhanced Sr optical clock according to the following criteria: i) achieving the collective strong-coupling regime on both the closed $^1S_0-^3P_1$ transition and clock transition to generate atomic entangled states and superradiance; ii) making the atom-cavity coupled system in the so-called ``bad cavity'' regime, i.e. $\kappa>\Gamma_{^3P_1}$; iii) maximizing the light-atom coupling homogeneity by trapping the atoms in a magic-wavelength lattice with commensurate spacing with respect to the probe cavity standing wave~\cite{Hosten16}. The system will then consist on two overlapped cavities, a Fabry-Perot cavity for the cQED system and a bow-tie ring cavity to obtain lattice confinement at the clock magic wavelength.

\begin{figure}[b]
\centering
\includegraphics[width=8.1cm,clip]{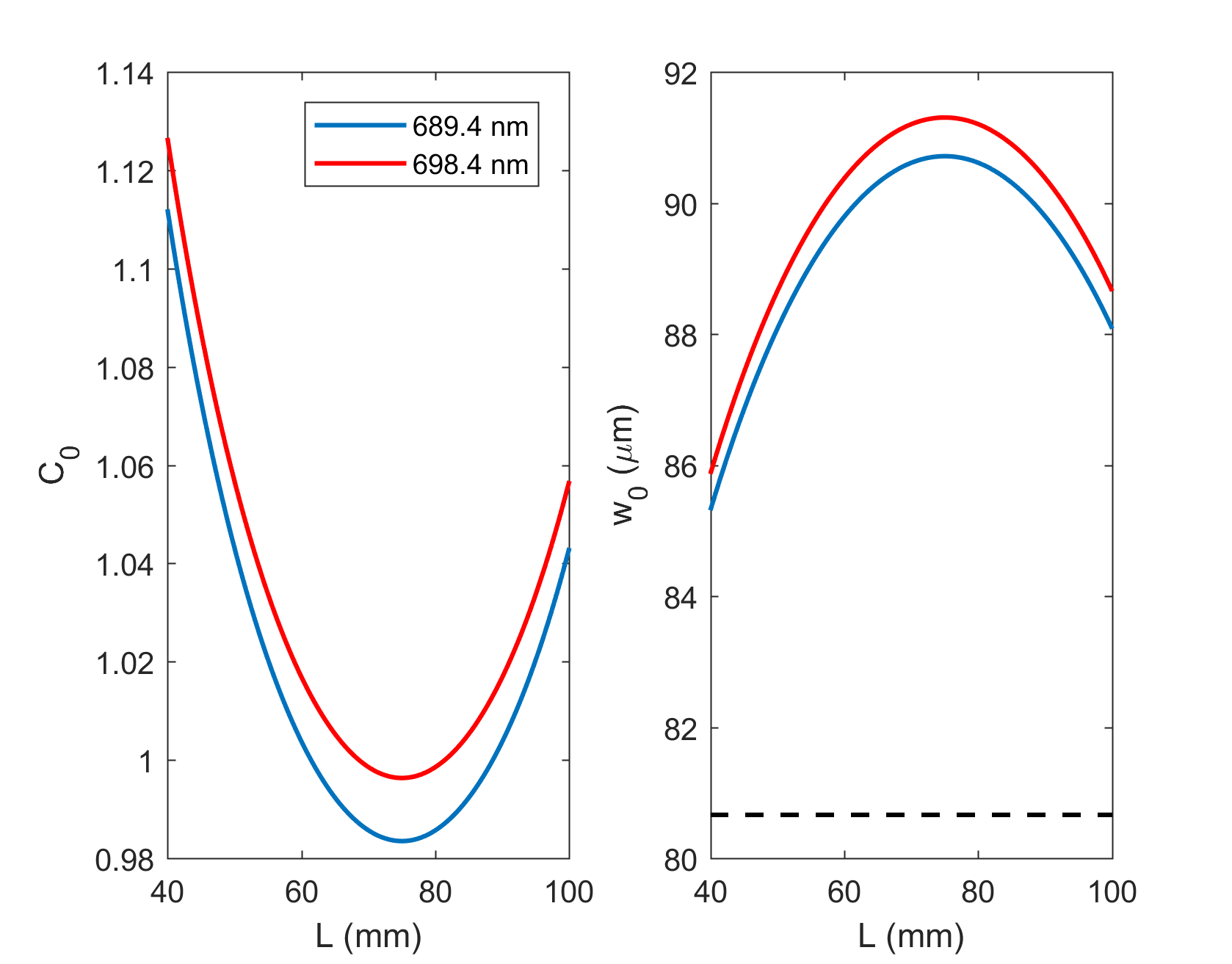}
\caption{Geometrical and coupling parameters of the strongly-coupled cavity at 694 nm. Left plot: Single-atom peak cooperativity as a function of the cavity length for a symmetric spherical cavity with $\mathcal{F} =$ 88000 at both wavelengths. Right plot: Cavity waist for both wavelengths as function of $L$, compared to the tangential waist of the ring lattice cavity (black dashed line).}
\label{fig:cqed_cavity}       
\end{figure}

\subsection{cQED system design}

For the cQED system, the main figure of merit in the atom-cavity coupling is the cavity cooperativity parameter $C_0 = 4g^2/(\Gamma\kappa)$ rather than the absolute cavity coupling strength $g$. The cavity cooperativity represents the ratio between the atom-light mode cross section and the cavity mode size multiplied by the finesse, and depends on the cavity optical parameters as follows~\cite{Tanji2011}: 

$$
C_0 = \frac{24 \mathcal{F}}{\pi k^2 w_0^2}.
$$

We see that the cavity cooperativity is not limited by the cavity length, as it happens for the coupling strength in typical cavity QED experiments~\cite{Walther2006}, so that using longer cavities makes easy to integrate the cQED system with magneto-optical trapping beams. In order to maximize the homogeneity of the light probing, cavity modes with larger radial size $w_0$ are desirable, but because of the inverse square dependence of $C_0$, larger modes require higher cavity finesse. On the other hand, high values of the finesse, besides being impractical, reduces the cavity decay rate $\kappa$ which is bounded by $\Gamma$. All the above considerations brought us to design the cQED cavity starting from a nearly-confocal geometry and constraining the cavity finesse to be lower than $10^5$. Fixing mirrors' radius of curvature to 75 mm, we get near-unity cooperativity at $L=$ 66 mm and $\mathcal{F} =$ 88000, which is also stable under both mirror misalignment and radius-of-curvature asymmetries. Figure~\ref{fig:cqed_cavity} shows the cavity cooperativity parameter and the cavity waist as a function of the cavity length. We observe that the system is robust under big changes of the cavity geometry, and thus to tilt misalignment.

The cQED cavity is currently under production. It will be monolithic, its spacer made of a low thermal expansion material (Zerodur) which allows wide optical access to both lattice and cooling laser beams. At both ends of the spacer, the two low-losses mirrors will be attached to two piezoelectric transducers for independent tuning of both the cavity length and atom-cavity mode coupling due to the cavity nodes positioning with respect to the trapping optical lattice.

\subsection{Lattice bow-tie cavity design}

\begin{figure}[t]
\centering
\includegraphics[width=8.31cm,clip]{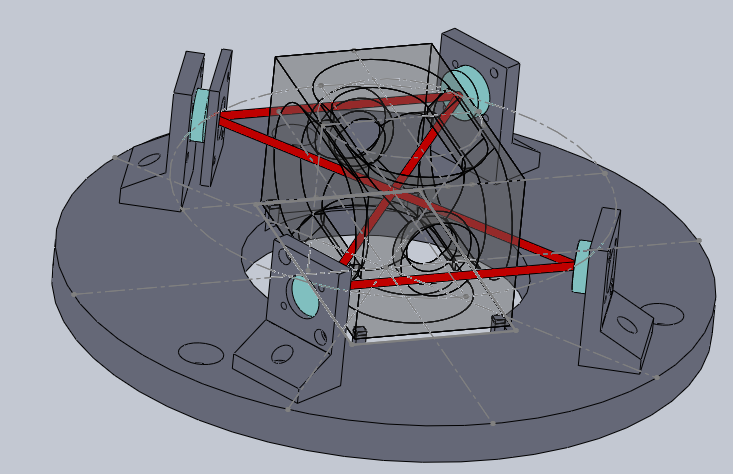}
\includegraphics[width=8.31cm,clip]{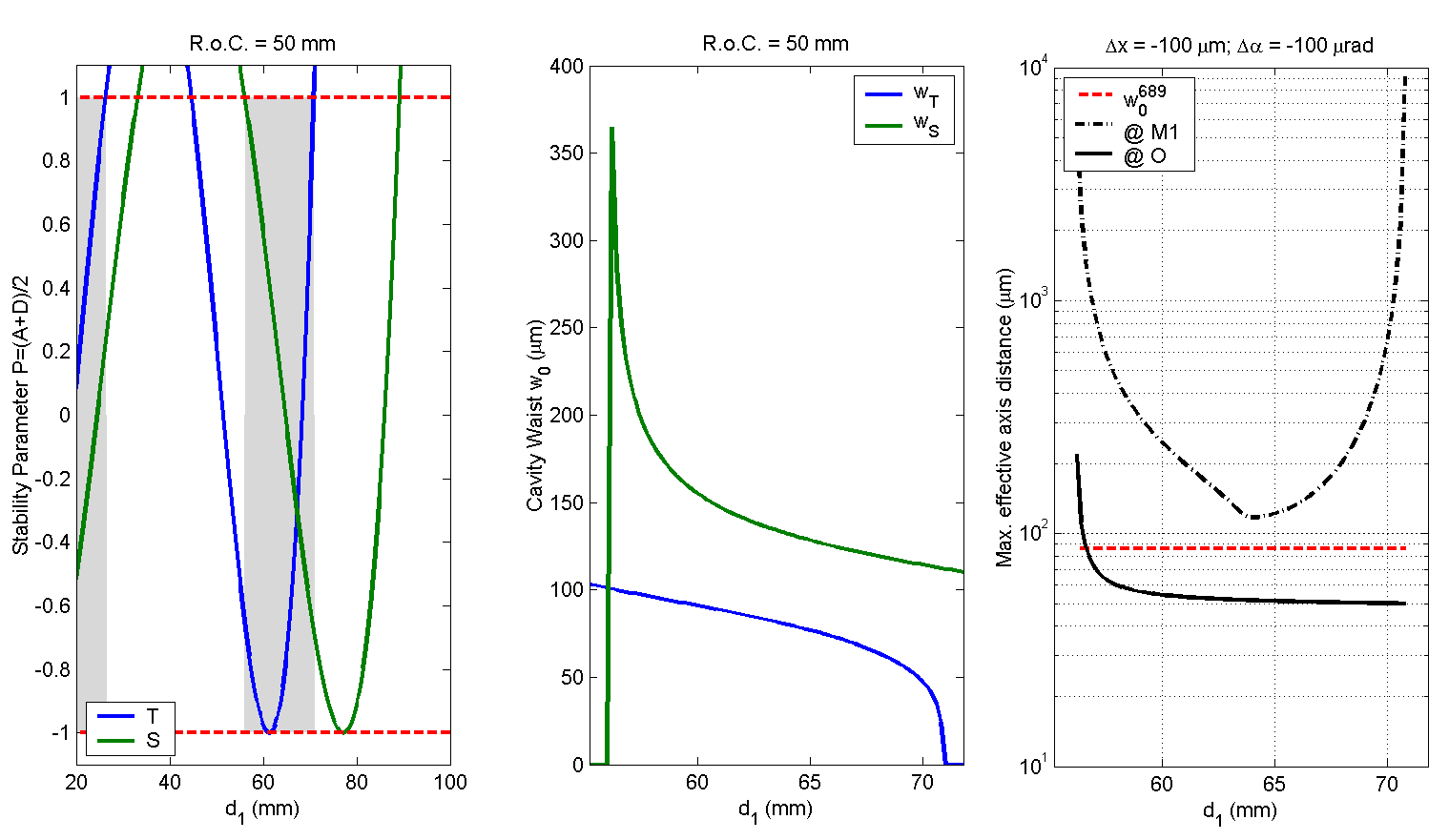}
\caption{Design of the lattice bow-tie cavity system. Upper panel: schematic drawing of the cavity-enhanced Sr optical clock at INRIM. The monolithic cQED cavity (transparent, only contour lines) is suspended in order to allow the bow-tie cavity beams (red) to intercept its optical axis at its center. Bow-tie cavity mirror mounts are just models. Lower panel: plots of the calculated stability region (left plot), cavity waist dimensions (center plot) and beam displacements for mirror misalignment on one mirror and at the cavity center (right plot).}
\label{fig:ring}       
\end{figure}

The lattice bow-tie ring cavity has been designed to surround the cQED cavity and to fit on a base plate compatible with a DN200CF flange. A preliminary design of its mechanical implementation in our system is depicted in Fig.~\ref{fig:ring}. 

Because the atom trapping point corresponds to the crossing point, the cavity must be symmetrical with all spherical mirrors with the same radius of curvature, so that the crossing point corresponds to a waist of the cavity and thus suppressing waveform aberrations. Furthermore, in order to satisfy the design criterion (iii), the intersection angle $\theta$ is constrained by~\cite{Morsch2006}

$$
\cos(\theta/2) = \frac{\lambda_p}{2 \lambda_l} ,
$$
where $\lambda_p$ is the probing light and $\lambda_l$ is the lattice wavelength which we choose to be 813 nm. If we consider $\lambda_p$ = 689.4 nm, then $\theta = 107.7^{\circ}$. In the case of the clock transition at 698.4 nm the angle is slightly different and it is 108.8$^{\circ}$. The angle $\theta$ sets both the mirrors incidence angle and one of the two cavity lengths, so that only two degrees of freedom are left: the mirror radius of curvature $R$ and namely the short adjacent mirror distance $d_1$. The cavity design has been optimized in order to minimize the waist at the cavity crossing point and minimize the cavity sensitivity to mirrors' misalignment. We performed numerical calculations according to the ABCD formalism including off-axis beam propagation~\cite{Martinez89}. The main results are shown in Fig.~\ref{fig:ring} (lower panel). The cavity parameters which optimize our design requirements are $R$ = 50 mm and $d_1$ = 63 mm. The mirror reflectivity is calculated to reach a trap depth $U_0$ = 1000 $E_r$ for an input power of 1 W, which implies a cavity finesse $\mathcal{F}_\text{BTC}=$ 350.

The design of the mechanical implementation of this cavity, and the choice of the material, is currently under development and we plan to send the final drawings for machining before the end of the year.

\section{Current status of INRIM's Sr optical clock}\label{sec:4}

\begin{figure}[t]
\centering
\includegraphics[width=8.31cm,clip]{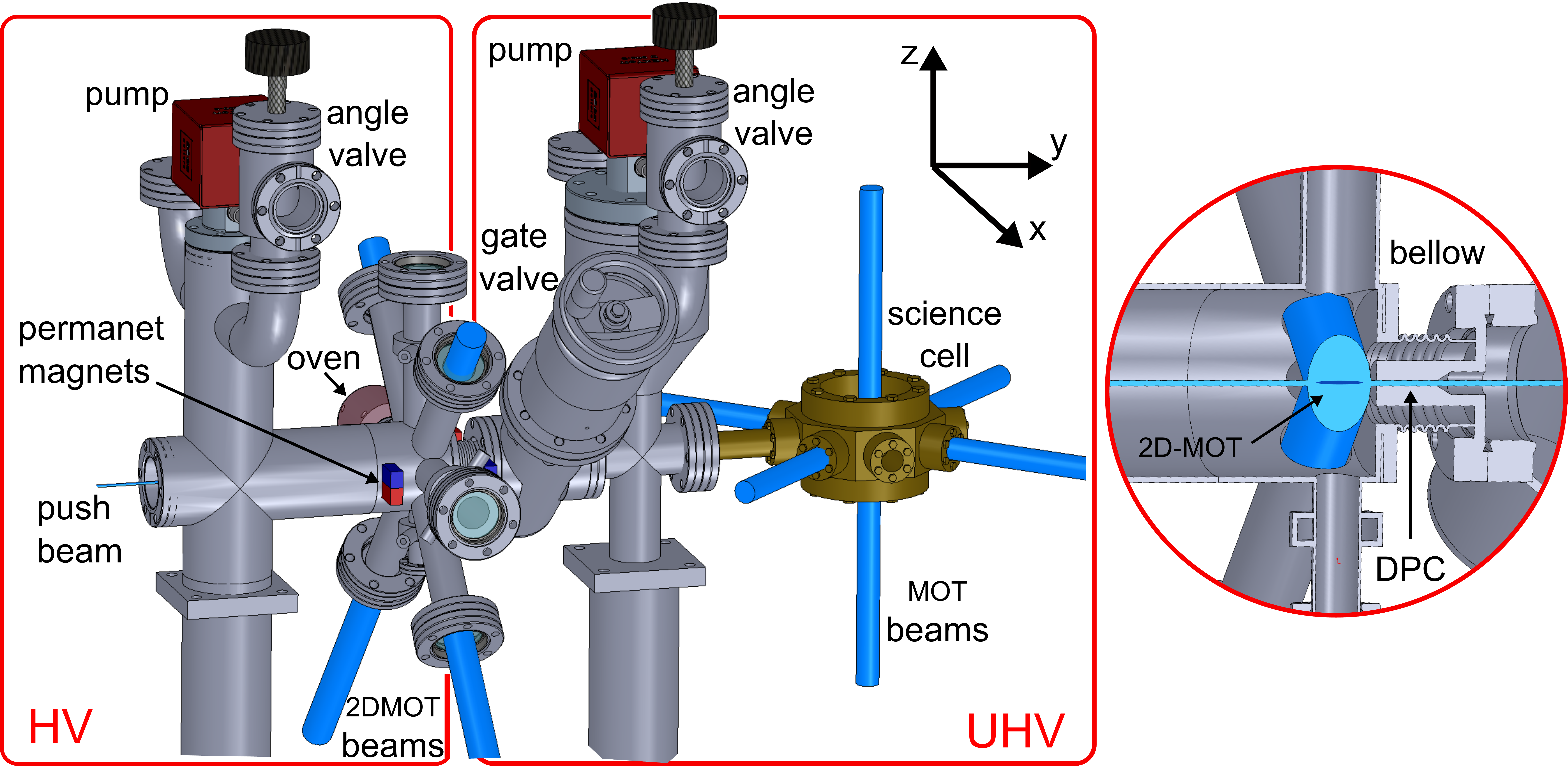}
\caption{Schematic drawing of the current vacuum system of the Sr optical clock at INRIM}
\label{fig-1}       
\end{figure}

The first building-block of our experiment has been completed. It consists on the realization of a sideband-enhanced cold atomic source of strontium atoms~\cite{barbiero2019}. A schematic drawing of the vacuum system and the corresponding important optical accesses are shown in Figure~\ref{fig-1}. In our setup, a two-dimensional magneto-optical trap (2D MOT) is transversely loaded from a collimated atomic beam and works as atomic funnel, delivering a cold atomic flux toward the desired spectroscopy region without exposing the latter to direct blackbody radiation from the oven, and suppressing hot atoms collisions when cooling lasers are switched off. In order to increase the capture efficiency of the 2D trap, we added a second frequency to the cooling beams, red-detuned with respect to the main cooling frequency. We performed a detailed study on enhancement and optimization of our cold strontium atomic beam flux by measuring the loading performances of the final 3D-MOT. Laser parameters of the 2D-MOT and sideband beams were scanned for optimal settings and compared with Monte Carlo simulations.

The final results about the maximum number of atoms in the MOT $N_{MOT}$ and the relative loading rate $L_{MOT}$, the total atomic flux $\Phi$ when the total laser power was 200 mW are summarized in Table~\ref{tab-1}. Compared to a conventional single-frequency 2D-MOT, we were able to obtain a 2.3 times enhancement of the atomic flux with a 50\% trapping efficiency.

\begin{table}[b]
\centering
\caption{Summary of the results obtained for our Sr cold atomic source in the case of single frequency 2D-MOT vs. the sideband-enhanced (SE) version~\cite{barbiero2019}}
\label{tab-1}       
\begin{tabular}{lll}
\hline
 & 2D-MOT & SE  \\\hline
$N_{MOT}$ & $5.3(2)\times 10^6$ & $1.25(4)\times 10^7$ \\
$L_{MOT}$ &$3.1(9)\times 10^8$ s$^{-1}$& $7.3(9)\times 10^8$ s$^{-1}$ \\
$\Phi$ & $6(1)\times 10^8$ s$^{-1}$ &$1.5(2)\times 10^9$ s$^{-1}$\\\hline
\end{tabular}
\end{table}

While the second stage cooling and trapping on the narrow linewidth intercombination transition at 689 nm is under development, we have realized a narrow ECDL at 1396 nm for clock spectroscopy after second-harmonic generation, and low-loss fiber network dissemination and single-branch stability transfer with a Ytterbium optical standard~\cite{Barbieri_2019}. Its estimated spectral linewidth is lower than 160 kHz but is prospected to be of the order of 10 kHz, greatly simplifying the frequency stabilization system to ultra-stable frequency references.

Finally, a new science chamber will be installed in order to house the new cavity-enhanced system. The vacuum chamber will be a titanium octagon with conflat flanges, two wide apertures to sit the cavity system (about 210 mm diameter), and on the sides 4 large optical windows (DN63CF) and other 4 smaller apertures (DN40CF). We plan to have the completed cavity-enhanced OLC in INRIM operational in 2020.

\section{Outlook}

In this brief article we have described how quantum technology can be employed for enhancing the performances of optical atomic clocks. In particular, how a cavity QED experiment can be integrated into an optical lattice clock experiment and the potential stability improvements, both in terms of projection noise reduction by entanglement in collective states, and by direct generation of lower-noise optical radiation through superradiance~\cite{bohnet2012}. 

The strontium OLC at INRIM is currently under construction. A cold atomic source with a novel trapping method based on two-frequency magneto-optical trapping has been demonstrated~\cite{barbiero2019}, while the design of a cavity-enhanced setup is almost complete. The generation of collective entangled states with reduced quantum uncertainty, and the study of superradiance in optical lattice are the next two milestones we aim to achieve toward a full quantum-limited optical clock system. 

\section*{Acknowledgments}
We thank C. D'Errico and L. Salvi for careful reading of the manuscript. We acknowledge funding of the project EMPIR-USOQS, EMPIR projects are co-funded by the European Union’s Horizon2020 research and innovation programme and the EMPIR Participating States. We also acknowledge QuantERA project Q-Clocks.

\bibliography{biblio}

%

\end{document}